\documentstyle[prl,aps,ulem]{revtex}
\begin{document}
\normalem
\twocolumn[\hsize\textwidth\columnwidth\hsize\csname
@twocolumnfalse\endcsname

\title{Temperature Dependence of the Flux Line Lattice Transition into Square
       Symmetry in Superconducting LuNi$_2$B$_2$C}

\author{M. R. Eskildsen$^{1,}$\cite{dpmc}, A. B. Abrahamsen$^1$,
        V. G. Kogan$^2$, P. L. Gammel$^3$, 
        K. Mortensen$^1$, N. H. Andersen$^1$ and P. C. Canfield$^2$}

\address{$^1$Ris\o \ National Laboratory, P. O. Box 49, DK-4000 Roskilde,
             Denmark \\
         $^2$Ames Laboratory and Department of Physics and Astronomy,
             Iowa State University, Ames, Iowa 50011 \\
         $^3$Bell Laboratories, Lucent Technologies,
             700 Mountain Ave., Murray Hill, New Jersey 07974}

\date{\today}
\maketitle

\begin{abstract}
We have investigated the temperature dependence of the
$\bbox{H} \parallel \bbox{c}$ flux line lattice structural phase transition
from square to hexagonal symmetry, in the tetragonal superconductor
LuNi$_2$B$_2$C ($T_{\text{c}} = 16.6$ K). At temperatures below 10 K the
transition onset field, $H_2(T)$, is only weakly temperature dependent. Above
10 K, $H_2(T)$ rises sharply, bending away from the upper critical field. This
contradicts theoretical predictions of $H_2(T)$ merging with the upper critical
field, and suggests that just below the $H_{\text{c2}}(T)$-curve the flux line
lattice might be hexagonal.
\end{abstract}

\vskip2pc] \narrowtext

Studies of the topology of the magnetic flux line lattice (FLL) in type-II
superconductors have a long history. Early neutron scattering experiments on
low-$\kappa$ superconductors such as niobium revealed a multitude of different
FLL symmetries and orientations, mainly determined by the symmetry of the 
atomic crystal structure in the plane perpendicular to the applied field
\cite{nb}. This is not surprising since deviations from the hexagonal FLL
characteristic of an isotropic superconductor, and the locking to the
crystalline lattice are driven by the symmetry of the screening current plane
and by nonlocal flux line interactions within a range determined by the
coherence length $\xi_0$, and the electronic mean free path $\ell$. Later,
similar effects were also observed in the strong type-II superconductor
V$_3$Si with $\kappa \approx 17$ \cite{christ85}, demonstrating that nonlocal
effects can be equally important in high-$\kappa$ materials.

Over the last couple of years, effects of nonlocality have been clearly
observed in the borocarbide superconductors. The borocarbides are quaternary
intermetallics with stochiometry {\em R}Ni$_2$B$_2$C ({\em R} = Y, Gd--Lu) and
a tetragonal unit cell (I4/{\em mmm}) \cite{canfie98}. These materials have
attracted attention due to the coexistence of superconductivity
({\em R} = Y, Dy--Tm, Lu) and antiferromagnetic ordering ({\em R} = Gd--Tm).
The borocarbides are strong type-II superconductors with Ginzburg-Landau (GL)
parameter, $\kappa =$ 6--15. The discovery of a square FLL in most of the
$\bbox{H} \parallel \bbox{c}$ phase diagram \cite{sqFLL}, which undergoes a
smooth transformation into hexagonal symmetry at fields below 1 kOe 
\cite{sq2hex}, was the first observation of a purely field driven FLL symmetry
transition.

Using nonlocal corrections to the London model and incorporating the symmetry
of the screening current plane obtained from band structure calculations, one
is able to calculate the FLL free energy, and thereby to determine the stable
FLL configuration in different fields in the borocarbides \cite{kogan}. The
model succesfully describes the nature of the FLL square to hexagonal symmetry
evolution in the borocarbides with $\bbox{H} \parallel \bbox{c}$ as the applied
field is reduced. Qualitatively, this can be understood as driven by the four
fold basal plane anisotropy which makes the vortex current paths ``squarish''
close to the core. At high densities this leads to a square FLL, whereas at
low fields the system appears isotropic resulting in a hexagonal FLL
\cite{afflec97}. The onset of the transition occurs as the field decreases,
commensing at a critical field $H_2$, determined by the range of the nonlocal
interactions \cite{kogan}. Later studies of the FLL in a set of samples doped
with various amounts of impurities to systematically change the nonlocality
range, showed a quantitative agreement between the predicted and measured onset
of the transition \cite{cheon98,gammel99}. The model also predicted a
reorientation transition of the rhombic FLL at a field $H_1 \ll H_2$, which
has been  subsequently observed \cite{mkpaul98,abraha99}.

Whereas there is a growing experimental and theoretical understanding of the
FLL geometry and transitions for fields well into the mixed state, there has
been little attempt to understand what happens at fields just below the upper
critical field. In this letter we report on detailed measurements of the
temperature dependence of the structural transition, $H_2(T)$, in
LuNi$_2$B$_2$C with the applied field parallel to the crystalline $c$-axis,
over a broad range of temperatures and applied fields. Below 10 K, $H_2(T)$ is
essentially constant. Above 10 K, $H_2(T)$ rises sharply, eventually bending
away from the $H_{\text{c2}}(T)$ curve, suggesting the existence of a hexagonal
FLL in the region just below the upper critical field.

The FLL in LuNi$_2$B$_2$C ($T_{\text{c}} = 16.6$ K) was studied using the small
angle neutron scattering spectrometer on the cold neutron beam line at the
Ris\o \ National Laboratory DR3 research reactor. The sample used was a 1 gram
single crystal, grown from a high temperature flux \cite{canfie98} using
isotopically enriched $^{11}$B to enhance the neutron transmission. Incident
neutrons with wavelengths in the range $\lambda_{\text{n}} = 6 - 15.6$ \AA \
and a bandwidth $\Delta\lambda_{\text{n}}/\lambda_{\text{n}} = 24$\% were used,
and the FLL diffraction pattern was collected by a position sensitive detector
at the end of a 6 m evacuated tank. Magnetic fields in the range $1.5$ to 10
kOe were applied parallel to the crystalline $c$-axis and to the incoming
neutrons. In order to keep the resolution of the spectrometer unchanged, the
neutron wavelength was varied in concert with the applied field to keep the
scattering angle, $2\theta = q_{10} \, \lambda_{\text{n}}/2 \pi$, constant.
Here $q_{10}$ is the FLL scattering vector shown in figure \ref{fig1}. The
measurements were performed at temperatures between $2$ and $20$ K, following
a constant field cooling procedure. A background signal obtained above
$T_{\text{c}}$ was subtracted from the data.

In figure \ref{fig1} we show three different FLL diffraction patterns obtained
in LuNi$_2$B$_2$C. In quasi-2D structures such as that of the FLL, the real
space flux line configuration and the reciprocal space diffraction pattern are
related by a simple $90^{\circ}$ rotation and relabeling of the axes. The
diffraction pattern therefore directly shows the FLL symmetry and orientation,
with the caveat that scattering from differently oriented domains is
superposed. For the square FLL there is no orientational degeneracy with
respect to the atomic lattice, due to the tetragonal crystal structure of the
borocarbides, as reflected in the top panel of figure \ref{fig1}. Here both
the FLL (1,0)- and (1,1)-reflections are clearly visible, and resolution
limited both in the radial and the azimuthal direction. The square FLL is
aligned with the scattering vector $q_{10}$ along the $[110]$ crystal
direction.

However, as soon as the FLL is distorted away from a square symmetry there is
a two fold degeneracy, since orientations with the long rhombic unit cell
diagonal along either the $a$- or the $b$-axis have the same energy.
Experimentally this will first manifest itself as an azimuthal broadening and,
when the distortion increases as a splitting of the (1,0)-reflections, with
the two peaks each belonging to a different domain orientation \cite{sq2hex}.
The magnitude of the splitting will eventually reach $30^{\circ}$ if the FLL
is transformed to a perfect hexagonal symmetry. The azimuthal broadening is
evident in the middle panel of figure \ref{fig1}, and in the bottom panel the
(1,0)-reflections are clearly split. The effect of the transition is also
reflected in the square FLL (1,1)-reflections. Ideally a radial splitting
should be observed, with one peak moving towards shorter $q$ and gaining
intensity and the other moving to longer $q$ and fading, to become respectively
hexagonal (1,0)- and (1,1)-reflections belonging to different domain
orientations. However, due to the very rapid decrease in the scattered
intensity with increasing $q$, one only observes a radial broadening and a
relative increase in the intensity.

In general, the azimuthal splitting was only directly observable when above
$\sim 10^{\circ}$, comparable to experimental resolution. To determine the
transition onset field $H_2(T)$, we analyzed the azimuthal intensity
distribution in each diffraction pattern at a scattering vector $q = q_{10}$,
where $q_{10} = \alpha \times 2 \pi (B/\phi_0)^{1/2}$, $\phi_0 = hc/2e$ is the
flux quantum and $\alpha$ is a constant which depends on the FLL symmetry: With
a FLL apex angle $\beta$ as shown in figure \ref{fig3},
$\alpha = (\sin \beta)^{-1/2}$, and hence $\alpha=1$  for the square case
($\beta = 90^{\circ}$) and $(2/\surd 3)^{1/2}= 1.075$ for the hexagonal FLL
($\beta = 60^{\circ}$ or $120^{\circ}$). In the top panel of figure \ref{fig2}
we show the temperature dependence of the azimuthal width for $H = 3$ kOe,
obtained by fitting to a single Gaussian. A clear broadening is seen for
$T > 10$ K. The low-$T$ width of $13.8^{\circ} \pm 0.2^{\circ}$ is a direct
measure of the azimuthal resolution, which was kept constant througout the
experiment. To determine the azimuthal splitting, the intensity distribution
at each temperature and field was then fitted to two Gaussians with a width
fixed of $13.8^{\circ}$, as demonstrated in the top panel inset. In the bottom
panel of figure \ref{fig2} the azimuthal splitting is plotted versus
temperature for a number of fields between $1.5$ and 6 kOe. At low temperatures
the splitting goes to $0^{\circ} \pm 2^{\circ}$ as the field is increased. As
$T$ is raised the splitting increases smoothly for $H \leq 2$ kOe, and abruptly
above a certain threshold for $H \geq 3$ kOe. The threshold is determined by
$H_2(T)$. In addition, one notes a crossing of the curves for $4.5$ and 6 kOe,
which means that $H_2(T)$ is a multivalued function of $T$.

Figure \ref{fig3} is a contour plot of the azimuthal splitting in a
$(H,T)$-phase diagram. Here we have used the $3^{\circ}$ contour, well above
the zero-splitting error of $2^{\circ}$, as the criterion for the transition
onset field. Below 10 K, $H_2(T)$ is nearly constant, equal to $2.25$ kOe. This
is somewhat higher than the litterature values \cite{sq2hex}, but as shown by
Gammel {\em et al.} \cite{gammel99}, $H_2$ depends strongly on the sample
purity as this affects the range of the nonlocal flux line interactions. Above
10 K, $H_2(T)$ increases sharply, bending away from $H_{\text{c2}}(T)$ and
thereby avoiding to cross the upper critical field. From this it is clear that
the square FLL is not stable at $H_{\text{c2}}$(T). The strong temperature
dependence contradicts predictions based on extended GL theory
\cite{dewild97,rosens99}, where $H_2(T)$ intercepts $H_{\text{c2}}(T)$ at a
finite field. An extrapolation of our measurements suggest that in the
vicinity of $H_{\text{c2}}(T)$, the FLL may become hexagonal.

Theoretically the nonlocal contribution to the vortex-vortex interaction
weakens when the temperature rises, and $H_2(T)$ should somewhat increase. This
relatively weak increase is essentially captured in the London approach, where
the order parameter $|\triangle|$ is taken as a constant in space
\cite{kogan98}. Moreover, $|\triangle|$ is assumed equal to the zero-field
uniform value $\triangle_0 (T)$ which determines the condensation energy
$F_{\text{N}}-F_{\text{S}}$ of uniform superconductors
($\propto \triangle_0^2$ at low temperatures and $\propto \triangle_0^4$ near
$T_{\text{c}}$). Being constant in space and field independent, the
condensation energy so defined is commonly disregarded in calculations of the
equilibrium FLL. Within this scheme, $|\triangle|$ affects only the value of
the penetration depth, $\lambda\propto 1/|\triangle|$. The model is good for
flux line spacings large relative to the core size $\xi(T)$, i.e. in small
fields, and for temperatures well below $T_{\text{c}}$.

However, with increasing density of vortices at a fixed $T$, the spatial
average $\langle |\triangle| \rangle$ decreases thus causing an overall
increase of the system energy \cite{hao91}. The $\langle |\triangle| \rangle$
suppression depends on the ratio $a/\xi(T)$, with $a$ being the intervortex
spacing. Therefore, at a given flux line density, $B/\phi_0$, the system energy
can be reduced if the intervortex distance is maximized. This can be
interpreted as an extra (to the London interaction) repulsion of vortices. This
repulsion is isotropic for cubic materials or for tetragonal crystals with
$\bbox{H} \parallel \bbox{c}$, and will lead to a deviation from square
symmetry and eventually to a hexagonal FLL (or triangular with the structure
determined only by the ratio $\xi_{\text{a}}/\xi_{\text{b}}$ for
$\bbox{H} \parallel \bbox{c}$). We speculate that this effect should become
dominant on approaching $H_{\text{c2}}(T)$ where $|\triangle|$ is strongly
suppressed, in accord with the data presented in this letter.

In addition, we have earlier reported  that in TmNi$_2$B$_2$C at low
temperatures the FLL is hexagonal near $H_{\text{c2}}(T)$. With decreasing
field, the FLL transforms first to a rhombic structure followed by the square
\cite{eskild98}. As $H$ is further reduced, the FLL should go through the same
evolution in inverse order, confirmed by low field decoration data
\cite{abraha99}. This material is antiferromagnetic below $1.5$ K, and it is
possible that the magnetic ordering has a detrimental effect on the nonlocality
range in fields near $H_{\text{c2}}(T)$. If so, both the antiferromagnetism and
the order parameter suppression would have similar effects on the FLL, which
may explain why the domain of ``reverse evolution'' is so broad in
TmNi$_2$B$_2$C. Finally, also in agreement with our hypothesis are data on the
FLL in V$_3$Si that show how a distorted triangular lattice which is seen at
$T < 5$ K and $H = 10$ kOe evolves toward hexagonal when $T$ approaches
$T_{\text{c}}(H)$ \cite{yethir99}. 

Within the microscopic theory applied to the mixed state, there is no
artificial separation of the electromagnetic (London) intervortex forces and
those due to the varying order parameter. This problem, however, has never been
addressed for the FLL \cite{klein87}. The GL approach does provide such a
complete description, but it is valid only near the zero-field $T_{\text{c0}}$.
The accuracy of results obtained within GL for $T < T_{\text{c0}}$ is difficult
to control. Moreover, the very existence of the GL expansion in powers of the
order parameter gradients away from $T_{\text{c0}}$ is questionable. The
coherence length $\xi(T)$ does not diverge as $T \rightarrow T_{\text{c}}(H)$;
as a consequence the gradients of $\triangle$ are not necessarily small, and
the convergence of the GL series is difficult (if not impossible) to monitor.
In this respect, calculations based upon keeping certain quartic terms in the
GL expansion while discarding others \cite{dewild97,rosens99} can only be
justified by a microscopic theory, which is still to be done.

In conclusion, we have in this letter reported measurement of the temperature
dependence of the FLL square to hexagonal symmetry transition in
LuNi$_2$B$_2$C. Our data show that the transition field $H_2(T)$, bends away
from the upper critical field line, $H_{\text{c2}}(T)$. This suggests that
while approaching the $H_{\text{c2}}(T)$ curve, the FLL is driven towards
hexagonal symmetry. We propose a possible qualitative argument for why this
might be the case. This poses new questions and challenges concerning the
phase diagram of the flux line solid.

This project has recieved support from the Danish Technical Research Council
and the Danish Energy Agency. M.R.E. has received support from the Christian
and Anny Wendelbo Foundation. A.B.A. is supported by the Danish Research
Academy. V.G.K. and P.C.C. are supported by the Director of Energy Research,
Office of Basic Energy Science under contract W-7405-Eng.-82.


\begin{figure}
\caption{FLL diffraction patterns from LuNi$_2$B$_2$C obtained at temperatures
         of 10 and 12 K and fields of $1.5$ and 3 kOe. The orientation relative
         to the crystallographic axes, and the FLL (1,0)-peak azimuthal width
         is shown in the top panel. Each diffraction pattern is a sum of 4
         measurements with the sample oriented to satisfy the Bragg condition
         for each of the main peaks/groups of peaks, and subtracted by
         background measurements taken at 20 K. The neutron wavelengths used
         were respectively $\lambda_{\text{n}} = 15.6$ \AA \ ($1.5$ kOe) and
         $11.0$ \AA \ (3 kOe). 
         The scattering close to $q = 0$ is an artifact of imperfect
         background subtraction.
         \label{fig1}}
\end{figure}

\begin{figure}
\caption{Top panel: The FLL (1,0)-reflection azimuthal width obtained by
         fitting a single Gaussian to the azimuthal intensity distribution, as
         a function of temperature at an applied field of 3 kOe.
         Inset: FLL azimuthal intensity distribution folded into one quadrant,
         at 3 kOe and 2 and 12 K. The 2 K data (offset for clarity) are well
         fitted by a single Gaussian of width $13.8^{\circ}$ equal to the
         experimental resolution. The 12 K data are fitted by two equal weight
         Gaussians with the same width, yielding a splitting of
         $10.5^{\circ} \pm 0.4^{\circ}$.
         Bottom panel: Azimuthal splitting versus temperature for applied
         fields between $1.5$ and 6 kOe. The error bars includes both the error
         on the direct fit of the splitting, and the uncertainty in the
         experimental resolution. The dashed lines are guides to the eye.
         \label{fig2}}
\end{figure}

\begin{figure}
\caption{$(H,T)$-phase diagram showing constant azimuthal splitting contours.
         The dashed line corresponds to a splitting of $3^{\circ}$, and is
         taken as the FLL symmetry transition onset field $H_2(T)$. Approaching
         $H_{\text{c2}}(T)$ the scattered intensity vanishes and the shaded
         area shows the range of our measurements. A schematic illustration of
         the FLL symmetry and apex angle is shown for the square (centre) and
         hexagonal case (low $H$, high $T$).
         \label{fig3}}
\end{figure}

\end{document}